\documentclass[prb,twocolumn,showpacs,showkeys,amssymb]{revtex4}

\usepackage{graphicx}% Include figure files
\usepackage{dcolumn}% Align table columns on decimal point
\usepackage{bm}% bold math

\begin{document}

%\tightenlines

\title{Inelastic light scattering and the excited states of 
many-electron quantum dots}
\author{Alain Delgado}
\affiliation{Centro de Aplicaciones Tecnol\'ogicas y
 Desarrollo Nuclear, Calle 30 No 502, Miramar, Ciudad Habana, Cuba}
\email{gran@ceaden.edu.cu}
\author{Augusto Gonzalez}
\affiliation{Instituto de Cibern\'etica, Matem\'atica y F\'{\i}sica, Calle 
 E 309, Vedado, Ciudad Habana, Cuba}
\email{agonzale@icmf.inf.cu}

\begin{abstract}
\bigskip
A consistent calculation of resonant inelastic (Raman) scattering
amplitudes for relatively large quantum dots, which takes account of 
valence-band mixing,
discrete character of the spectrum in intermediate and final states, and 
interference effects, is presented. Raman peaks in charge and spin
channels are compared with multipole strengths and with the density
of energy levels in final states. A qualitative comparison with the
available experimental results is given.
\end{abstract}


\maketitle

The inelastic (Raman) scattering of light by a semiconductor quantum dot
is an optical process which has proven to be very useful as a 
experimental technique to study excited states. \cite{Hawrylak,Heitmann}
The interpretation of a resonant Raman experiment requires, however, a big 
experimental and theoretical effort. The theoretical description is 
often so complicated that consistent calculations have been 
carried out only for the smallest dots. \cite{Eduardo,rusos}

In the present paper, we address the question about resonant Raman
scattering in a relatively large quantum dot, aimed at reproducing the
main features of the Raman phenomenology by means of a transparent and 
consistent computational scheme. Particularly, we focus on topics such as
the character (single-particle or collective) of the Raman peaks, the
role of interference effects, Raman peaks for spin-excited final states, 
the modifications of the spectrum as the background electron density is 
changed , or the evolution of the spectrum as the incident laser energy 
moves from close to the effective band gap to well above it. 
Results of calculations are presented for GaAs dots with AlGaAs barriers. 
A qualitative comparison with the available experimental results is also given.

Our starting point is the perturbation-theory expression

\begin{equation}
A_{fi}\sim \sum_{int} \frac{\langle f|H^+_{e-r}|int \rangle
\langle int|H^-_{e-r}|i \rangle}{h\nu_i-(E_{int}-E_i)+i\Gamma_{int}},
\label{eq1}
\end{equation}

\noindent
for the amplitude of Raman scattering. \cite{texto} The kets $|i\rangle$ and
$|f\rangle$ are written as: $|i\rangle=|\psi_i\rangle |N_i\rangle$,
$|f\rangle=|\psi_f\rangle |N_i-1,1_f\rangle$, where $|\psi_i\rangle$, and
$|\psi_f\rangle$ are initial and final $N$-electron states, $|N_i\rangle$
is a state with $N_i$ photons of frequency $\nu_i$, and 
$|N_i-1,1_f\rangle$ is a state with $N_i-1$ photons of frequency $\nu_i$ and 
one photon of frequency $\nu_f$. On the other hand, the 
intermediate states are written as: $|int\rangle=|\psi_{int}\rangle 
|N_i-1\rangle$,  where $|\psi_{int}\rangle$ contains, besides the initial 
$N$ electrons, an additional electron-hole pair. $H_{e-r}$ is the
electron-radiation interaction hamiltonian, and $\Gamma_{int} = 0.5$ meV 
-- a phenomenological broadening.

Eq. (\ref{eq1}) shows the difficulties in computing $A_{fi}$ for a dot
containing dozens of electrons. One should construct approximations to
$|\psi_f\rangle$ in a 30 meV excitation energy interval, in which there 
could be hundreds of states, and approximations to $|\psi_{int}\rangle$ in 
a 30 meV energy interval above the band gap. In the latter situation, 
hole-band mixing should be taken into account in order to describe scattering 
to spin-excited final states. Notice that interference effects may come out 
from the sum over intermediate states. The 30 meV upper bound in final 
states is a typical threshold for phonon excitations.

Commonly, one avoids computing the intermediate states by approximating 
the whole expression for $A_{fi}$. In Ref. [\onlinecite{Lipparini}],
for example, Raman intensities are almost identified with strength 
functions (modulated multipole strengths), in accordance to the 
interpretation given by authors of paper [\onlinecite{Heitmann}] of their 
results. This approximation to $A_{fi}$ neglects
contributions from single-particle final states and interference
effects from intermediate states. It is supposed to be valid for laser
energies well above the effective band gap. A second common 
approximation to $A_{fi}$, which also neglects interference effects, is
the so-called extreme resonance condition, in which $h\nu_i$ is very close
to the band gap. In this case, Raman intensities are almost identified with
excited-state luminescence peaks, i.e. with the peaks in the density
of final-state energy levels. This interpretation was used by the authors
of Ref. [\onlinecite{Hawrylak}].

In our calculations, we start from the exact quantum-mechanical expression
(\ref{eq1}), and construct Random-Phase approximations to 
$|\psi_f\rangle$, and Tamm-Dankoff approximations to $|\psi_{int}\rangle$. 
\cite{nuclear} Explicit formulae, which should be particularised to the pure 
electronic system, may be found in [\onlinecite{nuestro}]. As the number 
of electrons in the dot is supposed to be relatively high, we expect that 
the insertion of the mean field functions $|\psi_{int}\rangle$ and 
$|\psi_f\rangle$ into (\ref{eq1}) would lead to a qualitatively correct 
picture for the positions and intensities of Raman peaks. The 
Hartree-Fock (HF) basis is 
used throughout. For holes, the HF equations include the electron mean field 
and the heavy-light hole mixing, treated by means of the Kohn-Luttinger
hamiltonian. A typical calculation in a 42-electron dot involves 
around 60 many-particle final states (for a given multipolarity and spin), 
and around 2000 intermediate states.

\begin{figure}[ht]
\begin{center}
\includegraphics[width=.95\linewidth,angle=0]{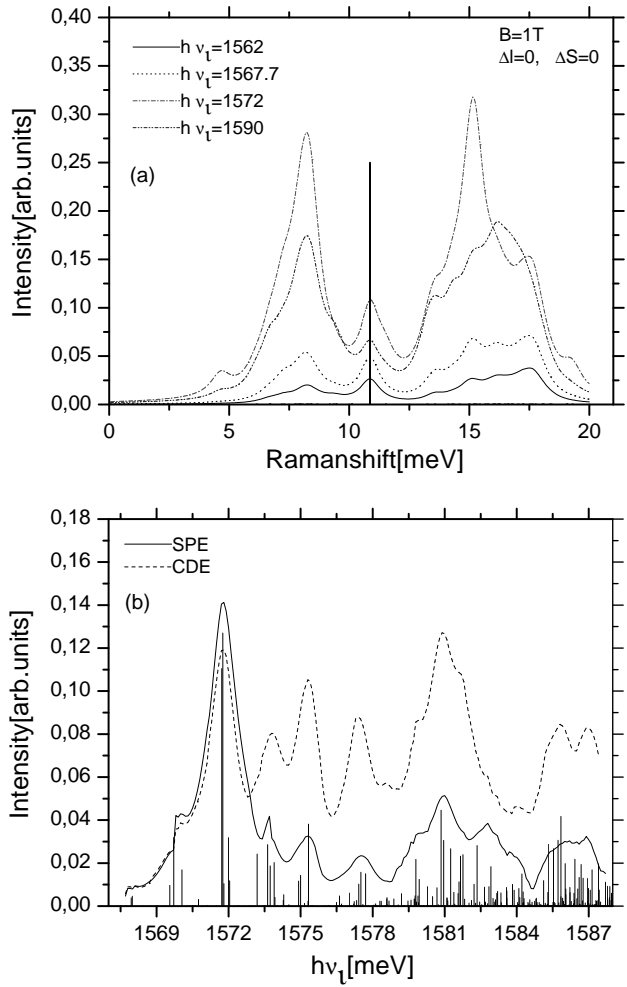}
\caption{\label{fig1} (a) Raman spectra for charge 
monopole final states at different frequencies of the incident light.
In the present and next figures, $h\nu_i$ is given in meV. A vertical line 
indicates the position of the collective state. (b) Raman intensities of 
one SPE and the collective state as functions of $h\nu_i$. The 
contribution of each intermediate state at resonance to the intensity of 
the SPE is also shown.} 
\end{center} 
\end{figure}

We first consider the extreme resonance condition, in which the incident
laser energy is very close the effective band gap. In Fig. \ref{fig1}(a), 
the calculated Raman intensities (squared amplitudes smeared out by means 
of Lorentzians of width $\Gamma_f= 0.5$ meV) for charge monopolar 
final states are shown. The dot is assumed to have a disk geometry with a 
25 nm width. The lateral confinement is parabolic, with 
$\hbar\omega_0=6$ meV, reproducing the observed position of the Kohn mode 
in the dots studied in Ref. [\onlinecite{Heitmann}]. The magnetic
field, equal to 1 Tesla, is perpendicular to the dot plane. The
final states considered in Fig. \ref{fig1} have the same spin and
angular momentum projection onto the magnetic field axis as the
initial state values. The nominal band gap is taken as 1560 meV. 
This gap is renormalized by Coulomb interactions to, approximately,
1567 meV. Raman amplitudes are computed in backscattering geometry. 
The initial and final
light polarization vectors are parallel. The laser energy is swept 
from 1562 to 1590 meV, that is from below the effective band gap
to 20 meV above it. The position of the collective monopolar state,
which carries more than 99 \% of the energy-weighted sum rule, is
signalized by a vertical line. In all of our calculations,
the incident (and scattered) light form an angle of 30 degrees with
the dot normal, which means that the maximum transferred momentum of 
light is $\Delta q_x\approx 0.8\times 10^{5} {\rm cm}^{-1}$. Under
these circumstances, Raman spectra are dominated by the lowest multipolar
final states.\cite{Heitmann}

Let us notice that the collective monopolar state is seen as a distinct 
peak at any laser frequency. However, to
final states carrying an almost zero fraction of the sum rule, for which
reason we will call them single-particle excitations (SPE), are
associated stronger, and wider, Raman peaks at energies below and above 
the collective state. 

Figure \ref{fig1}(b) shows the Raman efficiency of two particular final 
states, that is their Raman intensities as a function of $h\nu_i$. In 
general, the intensity of a given Raman peak in Fig. \ref{fig1}(a) is the 
result of three factors: (1) the number of final states contributing to 
it, (2) the Raman efficiencies of these states, and (3) interference 
effects.

Raman intensities corresponding to the collective monopolar state and to a 
SPE with Raman shift of 8.8 meV are 
shown in Fig. \ref{fig1}(b). The first interesting remark, in qualitative 
accordance with the existing observations, is that SPE are enhanced when 
$h\nu_i$ is close to the band gap, whereas collective states are enhanced 
as $h\nu_i$ is raised. On the other hand, the contribution of each 
particular intermediate state at resonance, i.e. $|\langle f|H^+_{e-r}|int 
\rangle \langle int|H^-_{e-r}|i \rangle|^2/\Gamma_{int}^2$, to the 
Raman intensity of the SPE is also included in the figure (vertical 
lines) in order to 
evaluate interference effects. Peaks in the Raman efficiency are related 
to particular intermediate states giving strong contributions to the sum 
(\ref{eq1}). In the figure, weak constructive or destructive interference 
in the neighborhood of these intermediate states can be appreciated.

\begin{figure}[hd]
\begin{center}
\includegraphics[width=.95\linewidth,angle=0]{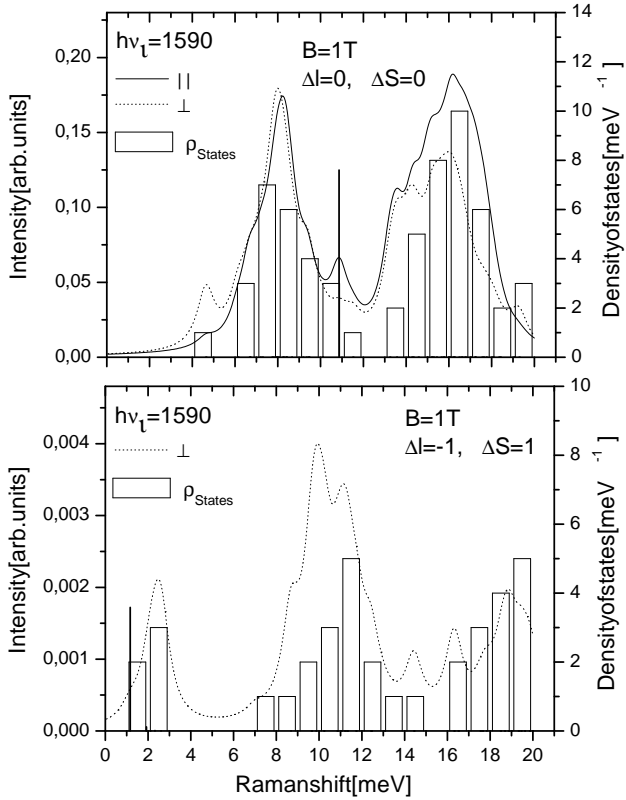}
\caption{\label{fig2} Upper panel: Raman intensities for charge monopole
final states. Lower panel: Raman spectra in spin dipole states. The
position of collective excitations are signalized by drop lines. 
The density of final-state energy levels is superposed in the figure.}
\end{center}
\end{figure}

The density of final state energy levels, computed from the 
Random Phase approximation to $|\psi_f\rangle$, is superposed
to the Raman spectra in Fig. \ref{fig2} in order to show its correlation 
with Raman peaks. The shape of the Raman spectrum depends on the 
frequency, but it is apparent that strong Raman peaks are associated to 
bunches of energy levels. These findings are in qualitative agreement with 
the experimental results of paper [\onlinecite{Hawrylak}]. 

In Fig. \ref{fig2}, curves labelled by a $\perp$ 
symbol represent a situation in which the incident an scattered light 
polarization vectors are orthogonal. For charge excitation (CE) channels, 
in the upper panel, the Raman spectrum in the 
orthogonal-polarization case is similar to that one in the 
parallel-polarization geometry. In the lower panel, Raman intensities in 
spin-excitation (SE) dipolar final states, are shown. By SE we mean states in 
which the total electronic spin projection is different from the initial 
state value. The difference in magnitude of peaks related to CE and 
SE in the orthogonal polarization would mean that SE peaks will be, in 
general, washed out, and only the lowest-energy SE levels, which are 
shifted to the left of CE states, have a chance to be measured. 
Notice also that the collective SE dipolar state, carrying more than 95 
\% of the sum rule, and which position is represented also by a drop 
line, is observed only as a very small shoulder in the Raman spectrum. 

\begin{figure}[ht]
\begin{center}
\includegraphics[width=.95\linewidth,angle=0]{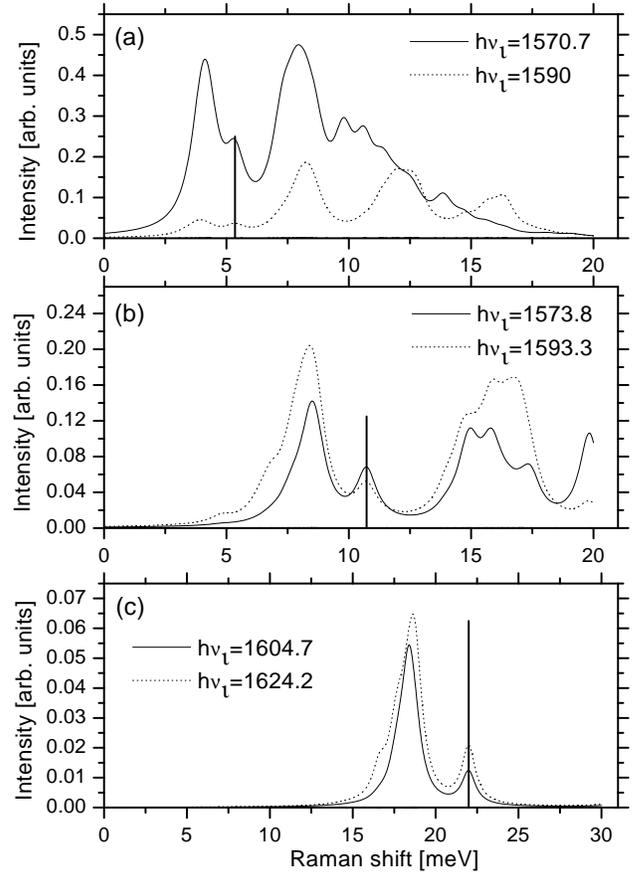}
\caption{\label{fig3} Raman spectra for charge monopole final states
 and parallel polarization. The magnetic field is set to zero. (a)
 $\hbar\omega_0=3$ meV, (b) $\hbar\omega_0=6$ meV, and (c) 
 $\hbar\omega_0=12$ meV.}
\end{center}
\end{figure}

Next, we consider the question about the effect of the 
density of the electronic cloud on the Raman spectra. In our 42-electron 
dot, we can control the density by varying the confinement strength:
$\rho\sim N^{1/2} m_e \omega_0/\hbar$. Notice, for instance, that the 
density of larger dots with around 200 electrons, as those studied in Ref. 
\onlinecite{Heitmann}, is similar to the density of our 42-electron dot 
when the parameter $\hbar\omega_0$ is doubled from 6 to 12 meV. 
Calculations were done also for a smaller frequency, $\hbar\omega_0=3$ 
meV, with the purpose of obtaining the whole picture. In our calculations, 
we  fixed the Coulomb-to-oscillator ratio of characteristic energies, 
given by the parameter $e^2m_e^{1/2}/(\kappa\hbar^{3/2}\omega_0^{1/2})$. 
\cite{Pade} It means that the relative strength of Coulomb interactions is 
kept constant when the density is varied.

The results are shown in Fig. \ref{fig3}. For each 
value of $\hbar\omega_0$, Raman spectra are computed for laser energies 5 
and 25 meV above the effective band gap. The curves can be qualitatively 
understood on simple grounds. An increase in $\omega_0$ leads to a scaling 
of energies. For example, in Fig. \ref{fig3} (c), only the first SP and 
the first collective peaks are seen, the rest of the spectrum is moved to 
higher energies. In accordance to this scaling, the density of energy 
levels decreases both in intermediate and final states. Thus, the 
intensity of the Raman peaks should decrease by roughly a factor of 4 as 
$\omega_0$ is doubled. The relative intensity of peaks depends on the 
Raman efficiency, as mentioned above.

\begin{figure}[ht]
\begin{center}
\includegraphics[width=.7\linewidth,angle=-90]{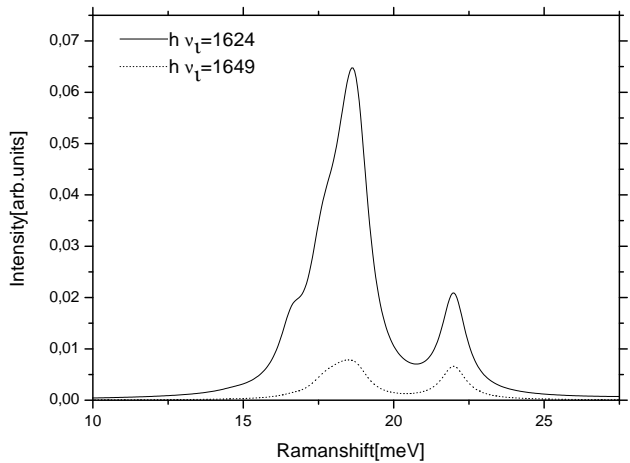}
\caption{Raman spectra for laser energies 25 and 50 meV above the 
effective band gap. Only the off-resonance contribution to the latter 
is shown. See explanation in the main text. \label{fig4}} 
\end{center}
\end{figure}

Finally, we want to discuss the situation in which the incident laser 
energy is well (around 50 meV) above the effective band gap.
\cite{Heitmann,Heitmann2} This regime is characterized by the following 
properties: (a) an overall decrease of Raman amplitudes, (b) a 
reinforcement of the peaks associated to collective states, and (c) a 
suppression of the SP peaks, except at the lower edge of SP excitations 
\cite{Sarma}.

In this case, it seems useful to distinguish the resonant and 
off-resonance contribution to the sum (\ref{eq1}). Only the latter
can be evaluated within our computational scheme because, for excitation 
energies around 50 meV, the intermediate states are doubtfully described 
by a simple Tamm-Dankoff approximation, and a constant $\Gamma_{int}$ is 
not a reasonable approximation. Indeed, one expects, for example, an 
increasing $\Gamma_{int}$ as we move to intermediate states with higher 
excitation energies. 

In Fig. \ref{fig4}, the off-resonance contribution to the Raman spectrum 
for laser energy 50 meV above the band gap is drawn, along with one 
spectrum, already shown in Fig. \ref{fig3} (c), corresponding to an 
incident photon energy 25 meV above the band gap. The off-resonance curve 
is computed from a sum which includes, as before, intermediate states 
with excitation energies below 30 meV. The smaller amplitude of this
curve is due to the big energy denominators. It is apparent, however, 
that the SP peak is stronger suppressed than the collective one. 

In general, the amplitude of the resonant contribution to (\ref{eq1}) 
shall decrease as $h\nu_i$ rises. The reason is that both $\langle 
f|H^+ |int\rangle$ and $\langle int|H^- |i\rangle$ decrease, 
while $\Gamma_{int}$ increases in this case. The reinforcement of 
collective states could be the effect of resonances in the intermediate 
states, but this is a question that requires a further work.

In conclusion, we presented calculations for the amplitudes of resonant
Raman scattering in 42-electron GaAs quantum dots based on the exact 
perturbation-theory formula (\ref{eq1}). To our knowledge, the largest 
previous calculation \cite{calc} considered a 12-electron dot, 
only one valence hole sub-band (the heavy hole), and assumed a spin 
unpolarized HF ground state.

Features related to SP (dominant) and 
collective final-state excitations are apparent when the incident laser
energy is varied in a 20 meV energy interval above the effective band gap. 
These features may be correlated to bunches of final-state energy levels 
and to particular intermediate states giving strong contributions to the 
sum (\ref{eq1}). Weak constructive or destructive interference effects 
can be appreciated in this regime. The intensity of spin-excitation peaks 
is shown to be one or two orders of magnitude weaker than the intensity of 
charge-excitation peaks for these laser energies. It means that only the 
lowest-energy spin excited states have a chance to be measured. On the 
other hand, for $h\nu_i$ well above the band gap, the off-resonance 
contribution to the Raman spectrum shows a strong suppression of SP peaks.  
These results are in qualitative agreement with the observations. 
 
There are many interesting points still uncovered. For example, to clarify 
the properties of the intermediate states giving a strong contribution to 
(\ref{eq1}). In quantum wells and for Raman shifts above 30 meV, peaks in 
the Raman efficiency of collective SE are shown to correspond to the 
absorption or emission of photons of particular frequencies, which are 
identified in PLE as a series of ``excitonic'' states \cite{Danan}. These 
``resonances'' in intermediate states could be the reason of strong 
enhancement of collective SE in quantum dots for $h\nu_i$ well above the 
band gap. Indeed, SE collective peaks were observed in this regime 
\cite{Heitmann}. These, and other, uncovered aspects of Raman scattering 
in quantum dots indicate the need for more experimental and theoretical 
work on this subject.

\begin{acknowledgments}
The authors acknowledge useful discussions with A. Cabo, F. Comas, 
E. Menendez-Proupin, R. Perez, and C. Trallero-Giner.
\end{acknowledgments}

\end{document}